\documentclass[journal]{IEEEtran}
\usepackage{amsmath,amsfonts,amsthm,amssymb,bm}
\usepackage{algorithmic}
\usepackage{array}
\usepackage[caption=false,font=normalsize,labelfont=sf,textfont=sf]{subfig}
\usepackage{textcomp}
\usepackage{stfloats}
\usepackage{url}
\usepackage{verbatim}
\usepackage{graphicx}
\usepackage{bbding}
\usepackage{multirow}
\usepackage{cite, bbm, xcolor}
\usepackage{refcount}
\usepackage{flafter}
\usepackage{makecell}
\usepackage{diagbox}
\usepackage{booktabs}
\usepackage{algorithm}

\newtheorem{remark}{Remark}
\theoremstyle{definition}

\usepackage{float}
\def\BibTeX{{\rm B\kern-.05em{\sc i\kern-.025em b}\kern-.08em
    T\kern-.1667em\lower.7ex\hbox{E}\kern-.125emX}}
    \expandafter\def\expandafter\normalsize\expandafter{%
    \normalsize%
    \setlength\abovedisplayskip{4pt}%
    \setlength\belowdisplayskip{4pt}%
    \setlength\abovedisplayshortskip{2pt}%
    \setlength\belowdisplayshortskip{2pt}%
}
\allowdisplaybreaks[4]
\usepackage{balance}

\begin{document}
\title{Transmit Pinching-Antenna Systems (T-PASS): Connecting Wired to Wireless Communications}
 \author{Deqiao Gan, Chongjun Ouyang, Yuanwei Liu, and Xiaohu Ge\vspace{-10pt}
         \thanks{D. Gan and X. Ge are with the School of Electronic Information and Communications, Huazhong University of Science and Technology, Wuhan 430074, Hubei, China. (e-mail: gandeqiao@hust.edu.cn, xhge@mail.hust.edu.cn).}
         \thanks{C. Ouyang is with the School of Electronic Engineering and Computer Science, Queen Mary University of London, London E1 4NS, U.K. (email: c.ouyang@qmul.ac.uk).}
         \thanks{Y. Liu is with the Department of Electrical and Electronic Engineering, The University of Hong Kong, Hong Kong. (e-mail: yuanwei@hku.hk).}
         }
\maketitle
\begin{abstract}
    A transmit pinching-antenna system (T-PASS) framework is proposed, in which a single pinched waveguide is employed to jointly serve one wired user equipment (UE) and multiple wireless UEs. The signal radiated by the pinching antennas (PAs) is used to serve the wireless UEs, whereas the residual guided signal at the waveguide termination is used to serve the wired UE. To facilitate T-PASS transmission and mitigate inter-user interference, a hybrid non-orthogonal multiple access (NOMA) scheme is introduced. Wireless UEs are scheduled by time-division multiple access (TDMA), and, in each slot, the scheduled wireless UE is paired with the wired UE through power-domain NOMA. Within this framework, the PA positions, PA radiation coefficients, power allocation, and TDMA time-slot allocation are jointly optimized to maximize a weighted sum rate (WSR). i) For the two-user case with one wired UE and one wireless UE, the optimal PA position and successive interference cancellation (SIC) decoding order are derived. Closed-form optimal power allocation is obtained, and a near-optimal PA radiation coefficient is determined through a low-complexity one-dimensional search. ii) For the multiuser case with one wired UE and multiple wireless UEs, four protocols with different PA-position and PA-radiation configurations are proposed. For each protocol, a low-complexity element-wise alternating optimization algorithm is developed to optimize the PA positions and radiation coefficients, while closed-form solutions are derived for the optimal power allocation and time-slot allocation. Numerical results are presented to show that: i) under typical T-PASS configurations, the wired UE is selected as the strong user in the optimal SIC decoding order; ii) the proposed T-PASS framework achieves a significantly higher WSR than conventional wireless-only PASS; and iii) additional flexibility in PA radiation and PA positioning further improves the WSR across the proposed protocols.
\end{abstract} 


\section{Introduction}
Driven by the growing demand for high-capacity and low-latency wireless connectivity, high-frequency bands, such as the 7-24 GHz upper mid-bands, have been widely regarded as a key enabler of next-generation networks \cite{bazzi2025upper}. Despite their abundant bandwidth, high-frequency links suffer from severe path loss and strong sensitivity to blockage, which makes reliable coverage difficult to guarantee  \cite{bjornson2025enabling}. To mitigate these impairments, large-aperture arrays have been adopted to provide high beamforming gains that compensate for propagation loss and blockage-induced attenuation \cite{bjornson2025enabling}. However, large arrays incur substantial deployment and interconnection overhead. They also offer limited channel reconfigurability, which restricts their ability to adapt to dynamic environments. These limitations have motivated increasing interest in \emph{reconfigurable antennas} that can tune electromagnetic (EM) properties, including polarization, operating frequency, and radiation pattern \cite{heath2025tri,castellanos2025embracing}. Such architectures support flexible \emph{EM beamforming} and facilitate efficient channel manipulation while keeping radio-frequency (RF) hardware complexity and power consumption at a moderate level \cite{heath2025tri,castellanos2025embracing}.

Representative examples include fluid antennas \cite{kit2021fluid} and movable antennas \cite{zhu2024movable}. These architectures reshape the channel through adaptive radiating states or element positions, which exploits spatial diversity and mitigates small-scale fading. Such designs can deliver notable performance gains, yet their practical reconfiguration range is often limited by a finite physical aperture that spans only a few to several tens of wavelengths \cite{yuanwei2025pass}. This limitation reduces their ability to compensate for large-scale attenuation at high-frequency bands \cite{yuanwei2025pass}. In addition, line-of-sight (LoS) blockage remains a fundamental bottleneck in high-frequency links. Aperture-limited reconfiguration typically cannot establish a high-gain alternative path quickly after blockage occurs \cite{yuanwei2025pass}. Many of these implementations also involve high fabrication cost, and key array properties, such as the number of antennas, are difficult to modify after deployment. These considerations motivate new reconfigurable antenna architectures that support scalable apertures and low-cost reconfiguration.
\subsection{Pinching Antennas}
To overcome these limitations, NTT DOCOMO proposed the \emph{pinching-antenna system (PASS)} at Mobile World Congress Barcelona 2021 as a new reconfigurable-antenna technology \cite{2022NTTDOCOMO}. PASS uses a dielectric waveguide as the transmission medium. EM waves are radiated into free space by attaching small dielectric particles at designated locations along the waveguide \cite{2022NTTDOCOMO}. These particles are referred to as \emph{pinching antennas (PAs)} \cite{2022NTTDOCOMO}. Each PA can be activated or deactivated independently at an arbitrary position along the waveguide, which enables dynamic reconfiguration of the antenna array. This architecture supports a flexible and scalable deployment strategy, which we refer to as pinching beamforming \cite{yuanwei2025pass}. Unlike conventional reconfigurable-antenna systems, PASS can extend the waveguide to a large length at low cost. This feature allows PAs to be placed near user equipments (UEs) to establish robust LoS links, which significantly reduces large-scale path loss and blockage-induced attenuation. PASS also offers low installation and maintenance cost, since reconfiguration only requires adding, removing, or repositioning inexpensive PAs.

Owing to these advantages, PASS has been envisioned as a promising technology for future high-frequency communications and has attracted increasing attention \cite{yuanwei2025tutorial,xu2025generalized}. In \cite{ding2024pass}, the achievable rate of a PASS with a single PA serving a mobile UE was analyzed. Building on this result, \cite{Tyr2025los} studied the outage probability of a single-PA PASS while accounting for in-waveguide propagation loss. These works established theoretical evidence that PASS can outperform conventional fixed-antenna systems under practical high-frequency impairments. To further characterize the fundamental performance limits of PASS, recent studies have investigated the array gain \cite{ouyang2025array}, the channel capacity \cite{ouyang2025capacity}, and the blockage probability \cite{ding2025losblockage}. The above works primarily focused on communication-theoretic metrics. As a further step toward physically grounded modeling, \cite{wang2025pass} developed an EM-theory-based characterization of PASS and derived an energy-radiation model for a PA. This model quantifies the fraction of guided power that leaks into free space when the waveguide signal launched from the feed point reaches a PA. Collectively, these results indicate that PASS mitigates large-scale path loss and LoS blockage and can outperform traditional reconfigurable-antenna technologies, including fluid and movable antennas \cite{yuanwei2025tutorial,xu2025generalized}.

Building on this theoretical groundwork, a range of pinching beamforming algorithms have been developed to optimize the placement of active PAs along the waveguide in single-user settings \cite{xu2024rate} and in multiuser PASS under orthogonal multiple access (OMA) \cite{xie2025nomapass,zeng2025energy} and non-orthogonal multiple access (NOMA) \cite{wang2024antenna,tegos2024minimum,zeng2025energyuplink,zeng2025sum}. Joint baseband and pinching beamforming designs were later investigated to further improve performance \cite{bereyhi2025downlink,gan2025NOMAPASS,sun2025multiuser}. In parallel, channel estimation \cite{xiao2025channel} and beam training \cite{lv2025beam,gan2025llm} techniques have been studied to acquire accurate channel state information (CSI) for efficient PA configuration. Learning-based approaches have also been explored to accelerate beamforming inference and PA control \cite{jia2025GNNPASS,jia2025PASSISAC}. Beyond communications-oriented studies, PASS has been applied to other emerging scenarios. PASS-aided over-the-air computation was investigated to reduce aggregation error in \cite{lyu2025aircom}. PASS-based designs for low-altitude networks were considered to improve throughput in \cite{guo2025learning}. PASS-assisted wireless power transfer was proposed in \cite{gan2025wpt} to balance energy delivery and communication performance. The application of PASS to physical-layer security was further examined in \cite{wang2025pinchingkaidi,jiang2025pinching}. Collectively, these representative studies provide a foundation for understanding and analyzing PASS in high-frequency networks.

\begin{figure}[!t]
    \centering
    \includegraphics[width=0.25\textwidth]{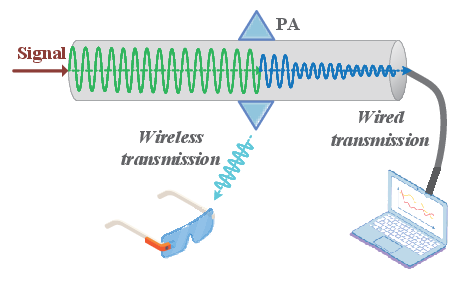}
    \caption{Illustration of the T-PASS.}
    \label{fig1}
    \vspace{-10pt}
\end{figure}

\subsection{Motivation and Contributions}
According to the EM model of PASS, each PA leaks a fraction of the guided wave for radiation. The remaining power continues to propagate along the dielectric waveguide and can be leaked by subsequent PAs \cite{wang2025pass}. In practice, the number of PAs on a waveguide is finite, and the per-PA radiation ratio can be limited by hardware constraints. A non-negligible residual forward wave can therefore remain after the last active PA. This residual guided signal must be properly handled at the waveguide termination. Otherwise, an impedance mismatch can induce reflections that generate a backward-travelling wave and standing-wave patterns, which distort the intended power distribution across PAs \cite{pozar2021microwave}. Most existing PASS studies adopt a travelling-wave abstraction. Under this abstraction, a matched termination absorbs the residual guided signal so that reflections are negligible and the residual power does not contribute to communications \cite{yuanwei2025tutorial}. In contrast, the residual power can be moderate when the number of PAs is limited or the radiation ratio is small, which creates an opportunity to recycle it as an additional information-bearing resource.

Motivated by this observation, we propose a \emph{power-recycling termination} in which the waveguide end is connected to a matched waveguide-to-receiver coupler that serves a dedicated wired UE \cite{pozar2021microwave}. The wired UE can represent, for example, an information kiosk in a library or a ticketing machine in a mall. The proposed termination preserves the low-reflection travelling-wave condition, which ensures that the backward-propagating wave remains negligible, while converting the residual guided power into a useful wired information stream. This design supports simultaneous wireless radiation through PAs and wired delivery at the terminal port, which improves the overall system utility. The resulting architecture, illustrated in {\figurename} {\ref{fig1}} and referred to as the \emph{transmit PASS (T-PASS)}, can simultaneously serve multiple wireless UEs and one dedicated wired UE.

Although the concept of T-PASS is intuitive, its design is not straightforward and cannot be obtained by a direct extension of existing wireless-only PASS schemes. Three challenges arise. First, a concise signal model that captures the wired communication link in PASS is currently unavailable. Second, a pinched waveguide is driven by a single radio-frequency (RF) chain, which limits the available spatial degree of freedom to one. Under this constraint, the simultaneous service of multiple wireless UEs and one wired UE becomes challenging, and a transmission framework that manages inter-user interference (IUI) in T-PASS remains to be developed. Third, the wired UE typically experiences a much stronger channel gain than the wireless UEs. This pronounced channel-quality disparity complicates the joint design of pinching beamforming and power allocation under heterogeneous quality-of-service (QoS) requirements.

Motivated by the potential of T-PASS and the above challenges, this article develops a signal model and a transmission framework for T-PASS that facilitate practical integration of wired and wireless services. The main contributions of this article are summarized as follows.
\begin{itemize}
    \item We propose a T-PASS architecture that enables simultaneous transmission to multiple wireless UEs and one dedicated wired UE through a single pinched waveguide. To control IUI, we develop a hybrid NOMA (H-NOMA) transmission framework. Time-division multiple access (TDMA) serves the wireless UEs without mutual interference, while power-domain NOMA pairs the wired UE with each scheduled wireless UE within its allocated time slot. This design exploits the residual guided power at the waveguide termination and improves the overall system throughput. Under this framework, we establish a signal model for the wired link and formulate a weighted sum-rate (WSR) maximization problem. The formulation jointly optimizes time-slot allocation, power allocation, successive interference cancellation (SIC) decoding order, PA placement, and PA radiation coefficients.   
    \item To obtain analytical insight, we study a two-user special case in which a single PA serves one wireless UE and one wired UE. We derive the optimal PA placement for this setting. On this basis, we show that under mild and practically relevant conditions, the wired UE should be treated as the strong user within the NOMA pair and should perform SIC decoding. We then derive a closed-form solution for the optimal power allocation. We also propose a low-complexity search method to optimize the PA radiation factor.
    \item We extend the design to a general multiuser setting with multiple PAs, multiple wireless UEs, and one wired UE. Guided by the two-user analysis, we fix the SIC order in each TDMA slot by assigning the wired UE as the strong user. We propose four H-NOMA-based T-PASS protocols, depending on whether the PA radiation coefficients and PA placements are shared across TDMA slots: \emph{fixed-radiation fixed-position (FR-FP)}, \emph{fixed-radiation adaptive-position (FR-AP)}, \emph{adaptive-radiation fixed-position (AR-FP)}, and \emph{adaptive-radiation adaptive-position (AR-AP)}. For each protocol, we derive closed-form solutions for the optimal time-resource allocation and power allocation. Based on these results, we develop efficient element-wise alternating-optimization algorithms to optimize the PA radiation coefficients and PA placements. 
    \item We provide numerical results to validate the advantages of the proposed framework and to demonstrate that: i) the proposed T-PASS achieves a significant weighted sum-rate gain compared with conventional wireless-only PASS in the multiuser setting; ii) proper design of PA radiation and PA placement further improves performance; and iii) among the four protocols, AR-AP provides the best performance, followed by AR-FP, FR-AP, and FR-FP.
\end{itemize}

The remainder of this paper is organized as follows. Section \ref{Section: System_Model} introduces the system model of T-PASS. Section \ref{Section: Two-User Case} studies the two-user case and develops the corresponding transmission design. Section \ref{Section: Two-User Case} extends the design to the multiuser case. Section \ref{Section: Numerical Results} presents numerical results and provides detailed discussions. Finally, Section \ref{Section: Conclusion} concludes the paper.
\subsubsection*{Notations}
Throughout this article, scalars and vectors are denoted by non-bold and bold letters, respectively. The transpose is denoted by $(\cdot)^{\mathsf{T}}$. The notations $\lvert a\rvert$ and $\lVert \mathbf{a} \rVert$ represent the magnitude of scalar $a$ and the norm of vector $\mathbf{a}$, respectively. The expectation operator ia denoted by ${\mathbbmss{E}}\{\cdot\}$, respectively. The sets $\mathbbmss{C}$ and $\mathbbmss{R}$ denote the complex and real spaces, respectively. The shorthand $[N]$ denotes the set $\{1,\ldots, N\}$. The ceiling operator is denoted by $\lceil\cdot\rceil$. The notation ${\mathcal{CN}}(\mu,\sigma^2)$ refers to a circularly symmetric complex Gaussian distribution with mean $\mu$ and variance $\sigma^2$. Finally, ${\mathcal{O}}(\cdot)$ denotes the standard big-O notation. 

\begin{figure}[!t]
    \centering
    \includegraphics[width=0.3\textwidth]{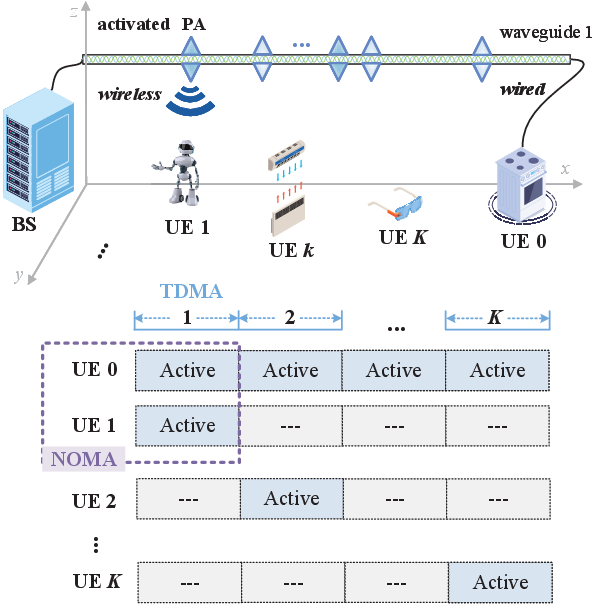}
    \caption{Illustration of the HNOMA-based transmission framework for T-PASS.}
    \label{fig3}
    \vspace{-10pt}
\end{figure}

\section{System Model}\label{Section: System_Model}
Consider a T-PASS communication network in which a base station (BS) employs a single pinched waveguide to serve $K$ single-antenna wireless UEs and one wired UE connected to the waveguide termination, as depicted in {\figurename} {\ref{fig3}}. The wireless UEs are distributed in a rectangular service region with side lengths $D_x$ and $D_y$ along the $x$- and $y$-axes, respectively. The position of wireless UE $k$ is denoted by ${\mathbf{u}}_k\triangleq[u_{k}^{x},u_k^{y},0]^{\mathsf{T}}$ for $k\in[K]$. The waveguide extends along the $x$-axis and spans the entire horizontal range of the service region.

Assume that $N$ PAs are deployed along the waveguide to radiate signals toward the wireless UEs. The location of PA $n$ is denoted by ${\bm\psi}_n=[\psi_n,0,d]^{\mathsf{T}}$ for $n\in[N]$, where $d$ is the deployment height of the waveguide. The PA positions satisfy $\psi_1<\psi_2<\ldots<\psi_N$. The feed point is located at the front-left end of the waveguide and injects the signal from the BS RF front end into the waveguide. Its location is ${\bm\psi}_0\triangleq[\psi_0,0,d]^{\mathsf{T}}$, with $\psi_0\leq \psi_{n}$ for all $n\in[N]$. The wired UE is connected to the waveguide termination through a waveguide-to-receiver coupler that is impedance-matched to the guided mode, which makes the end reflection negligible \cite{pozar2021microwave}. The wired UE is located at ${\mathbf{u}}_0\triangleq[u_{0}^{x},0,d]^{\mathsf{T}}$, and $u_{0}^{x}=\psi_0+D_x$ holds by construction. For notational convenience, the wired UE is indexed as UE $0$.
\subsection{Transmission Model}
The waveguide serves the wireless UEs through radiation from the deployed PAs and serves the wired UE through the residual guided signal collected at the waveguide termination. Since the waveguide is driven by a single RF chain, the spatial degree of freedom in the considered system is one. This constraint makes IUI a key challenge in multiuser transmission. To address this issue, we adopt a H-NOMA framework \cite{liu2024road} to serve the $K+1$ UEs. The time axis is partitioned into $K$ orthogonal slots. Slot $k$ is assigned to wireless UE $k$, and the wired UE is served concurrently within the same slot through power-domain NOMA, as shown in {\figurename} {\ref{fig3}}. The use of NOMA within each slot is motivated by two considerations. First, both UEs are driven by the same transmitted signal because the system has a single RF chain. Second, the guided link to the wired UE typically exhibits a much stronger channel gain than the radiated link to the wireless UE. This pronounced channel-quality disparity favors NOMA and supports efficient SIC-based reception \cite{liu2024road}, which improves throughput under mixed-service transmission with modest implementation complexity.
\subsection{Signal Model}
Let ${\mathbf{x}}\triangleq[x_1,\ldots,x_K]^{\mathsf{T}}\in{\mathbbmss{C}}^{K\times1}$ denote the transmit signal across the $K$ time slots, where $x_k$ is the superposed signal intended for wireless UE $k$ and the wired UE (UE $0$) in slot $k$ for $k\in[K]$. The signal $x_k$ is given by
\begin{align}
x_k=\sqrt{{p}_{k,k}}s_{k,k}+\sqrt{{p}_{k,0}}s_{k,0},
\end{align}
where $s_{k,k}\sim{\mathcal{CN}}(0,1)$ and $s_{k,0}\sim{\mathcal{CN}}(0,1)$ denote the normalized information symbols for wireless UE $k$ and UE $0$, respectively. The signal $x_k$ is injected at the feed point and propagates along the waveguide. The incident guided signal arriving at the first active PA in slot $k$ can be expressed as follows:
\begin{subequations}
\begin{align}
x_k^{(1)}&=x_k10^{-\frac{\kappa}{20}\lVert{\bm\psi}_1(k)-{\bm\psi}_{0}\rVert}{{\rm{e}}^{-\frac{2\pi}{\lambda_{\rm{g}}}\lVert{\bm\psi}_1(k)-{\bm\psi}_{0}\rVert}}\\
&=x_k10^{-\frac{\kappa}{20}\lvert{\psi}_1(k)-{\psi}_{0}\rvert}{{\rm{e}}^{-\frac{2\pi}{\lambda_{\rm{g}}}\lvert{\psi}_1(k)-{\psi}_{0}\rvert}},
\end{align}
\end{subequations}
where ${\bm\psi}_n(k)\triangleq[\psi_n(k),0,d]^{\mathsf{T}}$ denotes the location of the $n$th active PA in slot $k$ for $n\in[N]$, $\lambda_{\rm{g}} = \frac{\lambda}{n_{\rm{eff}}}$ is the guided wavelength, $\lambda$ is the free-space wavelength, $n_{\rm{eff}}$ is the effective refractive index of the dielectric waveguide \cite{pozar2021microwave}. Moreover, $\kappa$ denotes the average attenuation factor along the dielectric waveguide in dB/m \cite{yeh2008kappa}. 

According to EM coupling theory \cite{wang2025pass}, when $x_k^{(1)}$ reaches the first PA, a fraction of the guided signal is coupled out for free-space radiation and the remaining part continues to propagate along the waveguide toward subsequent PAs. Let $\delta_1(k)\in[0,1]$ denote the radiation coefficient of the first PA in slot $k$. Then the radiated signal from this PA is $\sqrt{\delta_1(k)}x_k^{(1)}$, while the residual guided signal that propagates toward the next PA is $\sqrt{1-\delta_1(k)}x_k^{(1)}$. The coefficient $\delta_1(k)$ can be adjusted by controlling the coupling length of the PA and its separation from the waveguide \cite{yuanwei2025tutorial}. 

By accounting for PA radiation across all $N$ PAs, the received signal at wireless UE $k$ in slot $k$ is given by
\begin{equation}\label{Wireless_Received_Signal}
\begin{split}
y_{k,k}&=x_k\sum\nolimits_{n=1}^{N}h_{\rm{o}}({\mathbf{u}}_k,{\bm\psi}_n(k))h_{\rm{i}}({\bm\psi}_n(k),{\bm\psi}_{0})\\
&\times\sqrt{\delta_n(k)}\prod\nolimits_{n'=1}^{n-1}\sqrt{1-\delta_{n'}(k)}+n_{k,k},
\end{split}
\end{equation}
where $h_{\rm{o}}({\mathbf{u}}_k,{\bm\psi}_n(k))$ denotes the free-space channel coefficient from PA $n$ at ${\bm\psi}_n(k)$ to UE $k$, and $n_{k,k}\sim{\mathcal{CN}}(0,\sigma^2)$ denotes additive white Gaussian noise (AWGN) with variance $\sigma^2$. The in-waveguide propagation coefficient from the feed point to PA $n$ in slot $k$ is denoted by $h_{\rm{i}}({\bm\psi}_n(k),{\bm\psi}_{0})$ and takes the following form:
\begin{subequations}
\begin{align}
h_{\rm{i}}({\bm\psi}_n(k),{\bm\psi}_{0})&\triangleq 10^{-\frac{\kappa}{20}\lVert {\bm\psi}_n(k)-{\bm\psi}_{0}\rVert}{{\rm{e}}^{-\frac{2\pi}{\lambda_{\rm{g}}}\lVert{\bm\psi}_n(k)-{\bm\psi}_{0}\rVert}}\\
&=x_k10^{-\frac{\kappa}{20}\lvert{\psi}_n(k)-{\psi}_{0}\rvert}{{\rm{e}}^{-\frac{2\pi}{\lambda_{\rm{g}}}\lvert{\psi}_n(k)-{\psi}_{0}\rvert}}.
\end{align}
\end{subequations}
The signal received by UE $0$ at the waveguide termination in slot $k$ can be written as follows:
\begin{equation}\label{Wired_Received_Signal}
\begin{split}
y_{k,0}=x_k \kappa_{\rm{c}}h_{\rm{i}}({\mathbf{u}}_0,{\bm\psi}_{0})\prod\nolimits_{n=1}^{N}\sqrt{1-\delta_{n}(k)}+n_{k,0},
\end{split}
\end{equation}
where $n_{k,0}\sim{\mathcal{CN}}(0,\sigma^2)$ denotes AWGN. The parameter $\kappa_{\rm{c}}\in(0,1]$ captures the coupling and insertion loss introduced by the impedance-matched waveguide-to-receiver coupler at the termination. For compactness, define the effective channels of wireless UE $k$ and UE $0$ in slot $k$ as follows:
\begin{align}
h_{k,k}&\triangleq\sum\nolimits_{n=1}^{N}h_{\rm{o}}({\mathbf{u}}_k,{\bm\psi}_n(k))h_{\rm{i}}({\bm\psi}_n(k),{\bm\psi}_{0})\nonumber\\
&\times\sqrt{\delta_n(k)}\prod\nolimits_{n'=1}^{n-1}\sqrt{1-\delta_{n'}(k)},\\
h_{k,0}&\triangleq\kappa_{\rm{c}}h_{\rm{i}}({\mathbf{u}}_0,{\bm\psi}_{0})\prod\nolimits_{n=1}^{N}\sqrt{1-\delta_{n}(k)}.\label{Wired_Channel_Model}
\end{align}
With these definitions, the received signals in \eqref{Wireless_Received_Signal} and \eqref{Wired_Received_Signal} can be written as $y_{k,k}=h_{k,k}x_k+n_{k,k}$ and $y_{k,0}=h_{k,0}x_k+n_{k,0}$, respectively.
\subsection{Channel Model}
PASS targets high-frequency bands \cite{2022NTTDOCOMO}, where LoS propagation typically dominates \cite{ouyang2024primer}. We therefore adopt a free-space LoS model for the spatial channel coefficient between the PA and the user \cite{ouyang2024primer}:
\begin{align}
h_{\rm{o}}({\mathbf{u}}_k,{\bm\psi}_n(k))\triangleq
\frac{\eta^{\frac{1}{2}}{\rm{e}}^{-{\rm{j}}k_0\lVert{\mathbf{u}}_k-{\bm\psi}_n(k)\rVert}}{\lVert{\mathbf{u}}_k-{\bm\psi}_n(k)\rVert},
\end{align}
where $\eta\triangleq\frac{c^2}{16\pi^2f_{\rm{c}}^2}$, $f_{\rm{c}}$ is the carrier frequency, $c$ is the speed of light, and $k_0\triangleq\frac{2\pi}{\lambda}$ is the wavenumber.
\subsection{Discussion on SIC Decoding}
{\color{black}For notational convenience, let $\Omega_k(k')\in\{1,2\}$ denote the decoding order of UE $k'\in\{0,k\}$ within each slot $k\in[K]$, where $\Omega_k(k')=1$ indicates that UE $k'$ is decoded first and $\Omega_k(k')=2$ indicates that UE $k'$ is decoded second. When $\Omega_k(k')>\Omega_k(k'')$, UE $k'$ first decodes the signal intended for UE $k''$ and then decodes its own signal. During the decoding of UE $k''$, UE $k'$ treats its own signal as interference. Two SIC decoding orders are possible in each TDMA slot. In SIC Order I, the wireless UE is treated as the weak user and the wired UE is treated as the strong user. This order corresponds to $\Omega_k(0)=1$ and $\Omega_k(k)=2$. In SIC Order II, the wireless UE is treated as the strong user and the wired UE is treated as the weak user. This order corresponds to $\Omega_k(k)=1$ and $\Omega_k(0)=2$.
\subsubsection{SIC Order I}
In the proposed T-PASS, UE $0$ is served by the residual guided signal collected at the waveguide termination. This link experiences deterministic waveguide attenuation and a stable coupling gain. Wireless UE $k$ is served by radiated signals and is subject to free-space path loss. As a result, UE $0$ typically has a stronger effective channel gain than UE $k$, and SIC Order I is usually preferred. Under SIC Order I, UE 0 decodes the wireless signal $s_{k,k}$ first and then decodes its own signal. Wireless UE $k$ decodes $s_{k,k}$ while treating the wired signal $s_{k,0}$ as interference.
\subsubsection{SIC Order II}
SIC Order II can arise in regimes where the radiated link becomes stronger than the guided link. This situation can occur when the PA is placed very close to wireless UE $k$ or when a radiation-dominant operating regime is considered. In this case, wireless UE $k$ performs SIC and UE $0$ treats the wireless signal as interference.
\subsubsection{Achievable Rate}\label{Section:System Model:Achievable Rate}
Next, we characterize the achievable rates under a given SIC order $\Omega_k(\cdot)$. Consider two UEs $k'$ and $k''$ within slot $k$ with $\Omega_k(k')>\Omega_k(k'')$. UE $k'$ first decodes the signal intended for UE $k''$ while treating its own signal as interference. The corresponding signal-to-interference-and-noise ratio (SINR) is given by $\gamma_{k,k' \rightarrow k''}
    = \frac{|h_{k,k'}|^2 p_{k,k''}}{|h_{k,k'}|^2 p_{k,k'} + \sigma^2}$. After cancelling UE $k''$'s signal, UE $k'$ decodes its own signal with signal-to-noise ratio (SNR) $\gamma_{k,k'}
    = \frac{|h_{k,k'}|^2 p_{k,k'}}{\sigma^2}$, which yields the following achievable rate:
\begin{equation}
    R_{k,k'} = \log_2(1 + \gamma_{k,k'}).
    \label{eq:rate_wl_own}
\end{equation}
UE $k''$ decodes its own signal while treating UE $k'$'s signal as interference. The SINR at UE $k''$ is $\gamma_{k,k''}
    = \frac{|h_{k,k''}|^2 p_{k,k''}}{|h_{k,k''}|^2 p_{k,k'} + \sigma^2}$, and the achievable rate is given by
\begin{equation}
    R_{k,k''} = \log_2(1 + \gamma_{k,k''}).
    \label{rate_n0_weak}
\end{equation}
Successful SIC at UE $k'$ requires that UE $k'$ can decode UE $k''$'s message at least as reliably as UE $k''$ can decode it. This condition is written as follows:
\begin{equation}
    \label{SIC_II}
    R_{k,k''} \le \log_2(1+\gamma_{k,k' \rightarrow k''}).
\end{equation}
\subsection{Problem Formulation}
In the considered H-NOMA-based T-PASS framework, the wired UE $0$ and the $K$ wireless UEs have different roles in terms of service priority and QoS requirements. To capture these differences in a tractable manner, we adopt a WSR objective and introduce nonnegative weights $w_0$ and $w_k$ for UE $0$ and wireless UE $k$, respectively. A larger weight indicates a higher service priority \cite{yang2016weightNOMA}. In addition, TDMA scheduling is characterized by time-sharing factors $\{\tau_k\}_{k=1}^{K}$, where $\tau_k\in[0,1]$ denotes the fraction of the transmission time allocated to wireless UE $k$ over the scheduling horizon. The combination of $\{w_0,w_k\}$ and $\{\tau_k\}$ captures both user priorities and the time-sharing pattern across slots, and it provides a controllable trade-off between overall spectral efficiency and fairness among UE $0$ and the wireless UEs. 

Under this hybrid NOMA transmission, we maximize the WSR by jointly optimizing the PA positions, the PA radiation coefficients, the power allocation, the decoding order, and the time-sharing factors. The resulting optimization problem is formulated as follows:
\begin{subequations}
    \label{wsumrate_k}
    \begin{align}
    &\max_{\psi_{n}(\cdot),\delta_n(\cdot),\Omega_k(\cdot),p_{k,k},p_{k,0}}
            \sum\nolimits_{k=1}^{K}\tau_k(w_{0}R_{k,0} + w_{k}R_{k,k})
        \label{prob:WSR_multiuser} \\
        &\quad\qquad{\rm{s.t.}} ~  \tau_kR_{k,k} \ge R_1^{\min}, k\in[K],\label{c:R1min}\\
        &\quad\qquad\quad~~ \sum\nolimits_{k=1}^{K} \tau_k R_{k,0} \ge R_0^{\min},\label{c:R0min}\\
        &\quad\qquad\quad~~ \eqref{SIC_II}, \Omega_k(k')>\Omega_k(k''),k'\ne k'',
        \label{c:SIC_k}\\
        &\quad\qquad\quad~~ p_{k,k''} \ge p_{k,k'}, \Omega_k(k')>\Omega_k(k''),k'\ne k'',
        \label{c:power}\\
        &\quad\qquad\quad~~ p_{k,1} + p_{k,0} = P_{\max},
        \label{c:power_all}\\
        &\quad\qquad\quad~~\sum\nolimits_{k=1}^{K} \tau_k =1, 0 \le \tau_k \le 1,k\in[K],\label{c:time_all}\\
        & \quad\qquad\quad~~\psi_n(k) \in [0,D_x], k \in [K], n \in [N],
        \label{c:position}\\
        & \quad\qquad\quad~~|\psi_n(k)-\psi_{n'}(k)|\geq \Delta_{\min}, n\ne n',
        \label{c:position1}\\
        & \quad\qquad\quad~~\delta_n(k)\in[0,1], k \in [K], n \in [N].
        \label{c:delta}
    \end{align}
\end{subequations}
In \eqref{c:R1min}, $R_1^{\min}$ denotes the minimum long-term throughput requirement for each wireless UE. This constraint prevents the TDMA scheduler from assigning a vanishing time fraction to a wireless UE. In \eqref{c:R0min}, $R_0^{\min}$ denotes the minimum long-term throughput requirement for UE $0$. Constraint \eqref{c:SIC_k} enforces successful SIC under the selected decoding order $\Omega_k(\cdot)$ within each slot, which is equivalent to \eqref{SIC_II} under SIC Order I or SIC Order II. Constraint \eqref{c:power} specifies the power ordering required by power-domain NOMA, where the weak user is assigned no less power than the strong user. Constraint \eqref{c:power_all} limits the per-slot transmit power by $P_{\max}$. Constraint \eqref{c:time_all} specifies a valid TDMA time-sharing policy. Constraint \eqref{c:position} restricts PA locations to the service interval along the $x$-axis, while constraint \eqref{c:position} ensures a minimum inter-PA distance $\Delta_{\min}$ to mitigate EM multiple coupling. Constraint \eqref{c:delta} constrains the radiation coefficients and reflects guided-power conservation without energy amplification.}

{\color{black}Problem \eqref{wsumrate_k} is challenging to solve due to the non-convex objective and the strong coupling among the optimization variables. The decoding-order selection in each time slot introduces an additional combinatorial component, which makes the problem NP-hard and complicates the search for a globally optimal solution. To address these difficulties, we first study a simplified two-user setting with a single wireless UE, i.e., $K=1$. This special case provides insight into the SIC decoding-order design and reveals key structural properties of the joint optimization. We then leverage these insights to develop efficient methods for the general multiuser case.}
\section{Two-User Case}\label{Section: Two-User Case}
{\color{black}This section investigates the two-user setting consisting of one wireless UE and one wired UE, as illustrated in {\figurename} {\ref{fig2}}. For further simplification, only a single PA is activated along the waveguide, which gives $N=1$. We first derive the optimal PA position. We then analyze the optimal SIC decoding order. With the decoding order fixed, we decouple the remaining variables and maximize the sum rate under a given PA radiation coefficient by deriving globally optimal power-allocation solutions. Finally, we derive a lower bound on the PA radiation coefficient. This bound avoids degenerate resource allocation while preserving sum-rate maximization under QoS constraints.

\begin{figure}[!t]
    \centering
    \includegraphics[width=0.3\textwidth]{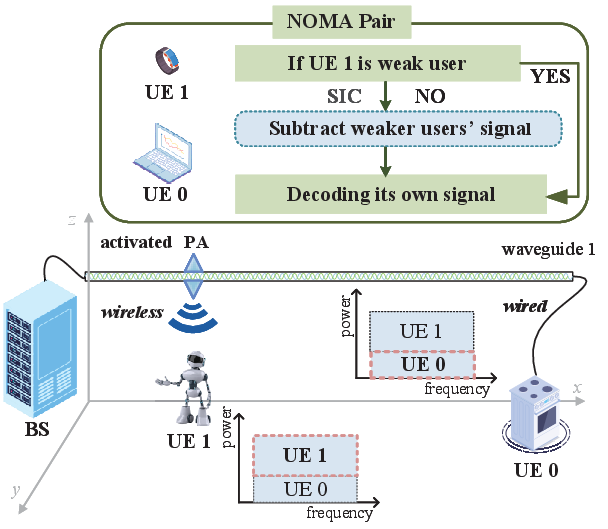}
    \caption{Two-user case design of the proposed T-PASS framework.}
    \label{fig2}
    \vspace{-10pt}
\end{figure}

\subsection{Problem Reformulation}
For brevity, the wired and wireless links are assigned the same service priority in this section. The goal is to reveal the fundamental performance gain of the proposed T-PASS framework. The WSR maximization in \eqref{wsumrate_k} therefore reduces to a sum-rate maximization problem by setting $w_0 = w_k = 1$. In this setting, only wireless UE $1$ is served through PA radiation, and only one PA is activated. For notational simplicity, $\psi_1(1)$ and $\delta_1(1)$ are denoted by $\psi_1$ and $\delta_1$, respectively. The resulting optimization problem is expressed as follows:
\begin{subequations}
    \label{sumrate_two}
    \begin{align}
        &\max_{{\psi}_{1},p_{1,1},p_{1,0},\delta_1,\Omega_1(\cdot)}
            R_{1,0} + R_{1,1}
        \label{prob:SR_2user_I} \\
        &~\quad\qquad{\rm{s.t.}} ~ R_{1,1} \ge R_1^{\min},R_{1,0} \ge R_0^{\min},\\
        &\quad\qquad\quad~~~\eqref{c:SIC_k},\eqref{c:power},p_{1,1}+p_{1,0}\leq P_{\max},\\
        &\quad\qquad\quad~~~\psi_1 \in [0,D_x],\delta_1\in[0,1].
        \label{c:twouser}
    \end{align}
\end{subequations}
Here, $p_{1,1}$ and $p_{1,0}$ denote the transmit powers allocated to wireless UE $1$ and UE $0$, respectively.
\subsection{Algorithm Design}
The variables in \eqref{sumrate_two} are tightly coupled. We therefore proceed in stages. We first determine the PA position and the SIC decoding order while treating the remaining variables as fixed. We then optimize the remaining variables based on these structural results.
\subsubsection{Optimization of the PA Position}
With $N=1$, the PA position only affects the wireless link. From \eqref{Wireless_Received_Signal}, the effective channel power gain of wireless UE $1$ is defined as $A_1\triangleq|h_{1,1}|^2$, which can be written as follows:
\begin{equation}
        \label{channelgain_11}
\begin{split}
        A_1 &= |h_{\rm{o}}(\mathbf{u}_1,\bm{\psi}_1)|^2 |h_{\rm{i}}(\bm{\psi}_1,\boldsymbol{\psi}_0)|^2 \delta_1 \\
        &=\frac{\eta\delta_1{\rm{e}}^{-2\alpha (\psi_1-\psi_0)}}{(\psi_1-u_1^x)^2+c_y}= \frac{\eta\delta_1}{D_{\rm{v}}^2}{\rm{e}}^{-2\alpha D_{\rm{i}}},
\end{split}
\end{equation}
where $\alpha \triangleq \frac{\kappa \ln 10}{20}$, $c_y\triangleq (u_1^y)^2+d^2$, ${\bm\psi}_1=[\psi_1,0,d]^{\mathsf{T}}$ denotes the PA location, $D_{\rm{v}} \triangleq \|\mathbf{u}_1-\boldsymbol{\psi}_1\|=\sqrt{(\psi_1-u_1^x)^2+c_y}$ denotes the distance between UE $1$ and the PA, and $D_{\rm{i}} \triangleq\psi_1 - \psi_0$ denotes the distance from the feed point to the PA along the waveguide. Referring to Section \ref{Section:System Model:Achievable Rate}, The achievable rate of UE $1$ increases monotonically with $A_1$ under either SIC order. Therefore, sum-rate maximization favors a PA position that maximizes $A_1$. This design is independent of the SIC order and of the power allocation.

Maximizing $A_1$ creates a trade-off between reducing the \emph{free-space path loss}, which favors placing the PA close to UE $1$, and reducing the \emph{in-waveguide attenuation}, which favors placing the PA close to the feed point. This trade-off becomes pronounced when UE $1$ lies near the service-region boundary. The optimal PA position is therefore obtained from the following:
\begin{align}
\psi_1^{\star}\triangleq\arg\max\nolimits_{\psi_1\in[0,D_x]}A_1.
\end{align}
The solution is given in closed form by \cite[Lemma 1]{xu2025passposition} as follows:
\begin{align}\label{SS_Optimal_Location}
{\psi}_{1}^{{\star}}=\left\{\begin{array}{ll}
\psi_{0}             & {c_y\geq\frac{1-(2\alpha \hat{u}_1^x-1)^2}{4\alpha^2}}\\
u_{1}^x+\frac{-1+\sqrt{1-4\alpha^2c_y}}{2\alpha}           & {\rm{Else}}
\end{array}\right..
\end{align}
where $\hat{u}_1^x\triangleq u_{1}^x-\psi_0$. Existing results show that, under typical PASS configurations, ${\psi}_{1}^{{\star}}$ satisfies \cite{xu2025passposition,ouyang2025uplink}
\begin{align}\label{SS_Approximated_Optimal_Location}
{\psi}_{1}^{{\star}}\approx u_1^x, 
\end{align}
which indicates that the optimal PA position nearly aligns with the projection of UE $1$ onto the waveguide.}
\subsubsection{Optimization of the SIC Order}
{\color{black}To determine the decoding order, we compare the effective channel power gains of the two users. By \eqref{SS_Optimal_Location} and \eqref{SS_Approximated_Optimal_Location}, the PA is placed close to the wireless UE to mitigate free-space path loss. The wireless-channel gain is maximized when the UE lies directly beneath the waveguide, i.e., $u_1^y=0$, and when the PA is aligned with the UE projection on the waveguide, i.e., $\psi_1=u_1^x$. Any lateral displacement increases the propagation distance and reduces the channel gain. Under this best-case geometry, the minimum distance between the PA and UE 1 equals the waveguide height $d$, and the wireless channel power gain satisfies
    \begin{equation}
        \label{channelgain_11}
        A_1 = \frac{\eta\delta_1}{d^2}{\rm{e}}^{-2\alpha D_{\rm{i}}}.
    \end{equation}
    Next, define the channel power gain of the wired link as $A_0 \triangleq |h_{1,0}|^2$. With $N=1$, the termination receives the residual guided signal, and it follows from \eqref{Wired_Channel_Model} that
    \begin{equation}
        \label{channelgain_10}
        A_0 = \kappa_{\rm{c}}^2 |h_{\rm{i}}(\mathbf{u}_0,\boldsymbol{\psi}_0)|^2 (1-\delta_1) = \kappa_{\rm{c}}^2 (1-\delta_1){\rm{e}}^{-2\alpha D_{x}}.
    \end{equation}

    To obtain a conservative comparison, we upper bound the wireless gain by setting $D_{\rm{i}} = 0$, which gives $A^{\star}_1 \triangleq \frac{\eta\delta_1}{d^2}\geq A_1$. The gain ratio then satisfies
    \begin{equation}
        \label{A0A1_maxratio}
        \frac{A_1^\star}{A_0} = \frac{\delta_1}{1-\delta_1}\left(\frac{\lambda}{d}\right)^2 \frac{{\rm{e}}^{2\alpha D_x}}{16\pi^2\kappa_{\rm{c}}^2}.
    \end{equation}
It is convenient to define $G_{\rm{ins}}\triangleq\left(\frac{\lambda}{d}\right)^2 \frac{{\rm{e}}^{2\alpha D_x}}{16\pi^2\kappa_{\rm{c}}^2}$, which is independent of $\delta_1$. According to existing tests and measurements, the impedance-matched waveguide-to-receiver coupler introduces coupling and insertion loss characterized by $\kappa_{\rm{c}} \in [{0.5},{0.9}]$ \cite{holloway2017fully,Zhan2025Electronics_PDW_Microstrip}. The considered PASS operates at centimeter-wave or millimeter-wave frequencies, and the term $\frac{\lambda}{d}$ is typically in the range $10^{-3}$--$10^{-2}$ in linear scale under representative geometries. Typical deployment lengths satisfy $D_x\in[10,100]$ m. For a representative dielectric waveguide, $\alpha=0.0092~{\text{m}}^{-1}$ has been reported in \cite{ding2024pass,xu2025passposition,ouyang2025uplink}, which yields ${\rm{e}}^{2\alpha D_x}\in[1.2,6.3]$. As a result, ensuring $\frac{A_1^\star}{A_0}\geq1$ requires $\delta_1\geq1-10^{-6}$. These values imply that the wireless link remains significantly weaker than the guided wired link unless $\delta_1$ is extremely close to one.
\vspace{-5pt}
\begin{remark}\label{remark_noma_1}
The above comparison uses a best-case upper bound on the wireless gain. Even under this favorable geometry, the wired link achieves a much larger channel gain than the wireless link under typical PASS parameters.
\end{remark}
\vspace{-5pt}
\vspace{-5pt}
\begin{remark}\label{remark_noma_2}
The radiation coefficient $\delta_1$ is governed by EM coupling between the PA and the waveguide. It depends on the interaction (coupling) length and device geometry. This dependence is sensitive to fabrication tolerances and environmental variations, such as temperature. As a result, $\delta_1$ cannot be tuned with arbitrarily high precision over $(0,1)$. The extreme regime required to overturn the guided-link advantage, such as $\delta_1\geq1-10^{-6}$, is not practically controllable.
\end{remark}
\vspace{-5pt}
These observations indicate that UE $0$ should be treated as the strong user in the NOMA pair and should perform SIC under typical PASS configurations. This conclusion follows from an upper bound on $A_1$, so it applies to all wireless UE locations within the service region. The decoding order is therefore fixed as follows:
    \begin{equation}
        \label{optimalSIC}
        \Omega_1(1)=1,\quad\Omega_1(0)=2.
    \end{equation}
    
Next, we consider an average-case setting in which the wireless UE is uniformly distributed over the service region, which implies that $D_{\rm{i}}$ is uniformly distributed on $[0,D_x]$. The average channel power gain of the wireless UE is given by
    \begin{equation}
        \label{aver_channelgain0}
        \begin{aligned}
            \hat{A}_1 & \triangleq \frac{1}{D_x}\int_{0}^{D_x}\delta_1 {\rm{e}}^{-2\alpha D_{\rm{i}}}|h_{1,1}|^2 {\rm{d}} D_{\rm{i}} \\
            & =\frac{\delta_1\eta}{D_x}\int_{0}^{D_x}\frac{{\rm{e}}^{-2\alpha D_{\rm{i}}}}{d^2} {\rm{d}} D_{\rm{i}}=
            \frac{\delta_1\eta}{d^2}\frac{1-{\rm{e}}^{-2\alpha D_x}}{2\alpha D_x}.
        \end{aligned}
    \end{equation}
It follows that
        \begin{equation}
            \label{A1A0_ratio}
            \begin{aligned}
                \frac{\hat{A}_1}{A_0} = \frac{\delta_1}{1-\delta_1}\left(\frac{\lambda}{d}\right)^2 \frac{1}{16\pi^2\kappa_{\rm{c}}^2}
                \frac{{\rm{e}}^{2\alpha D_x}-1}{2\alpha D_x}.
            \end{aligned}
        \end{equation}
This expression leads to the same conclusion as \eqref{A0A1_maxratio}. Under typical PASS parameters, $\delta_1$ must be driven to an extreme operating point that is impractical to control if the wireless link is required to exceed the guided wired link on average. Therefore, UE $0$ remains the strong user under both best-case and average-case channel conditions, and the SIC order can be fixed accordingly.}
\subsubsection{Optimization of the Power Allocation}
{\color{black}With the SIC order fixed as in \eqref{optimalSIC}, UE $0$ is the strong user and performs SIC. The achievable rates of UE $0$ and UE $1$ are
\begin{subequations}
\begin{align}
    R_{1,0} &= \log_2\left(1+\frac{A_0 \beta_1 P_{\max}}{\sigma^2}\right),\label{R_0_gain}\\
    R_{1,1} &= \log_2\left(1 + \frac{A_1 (1-\beta_1)P_{\max}}{A_1 \beta_1 P_{\max} + \sigma^2}\right),\label{R_1_gain}
\end{align}
\end{subequations}
where $\beta_1 \triangleq \frac{p_{1,0}}{P_{\max}}$ denotes the power-allocation coefficient for UE $0$, and $p_{1,1} = (1-\beta_1)P_{\max}$. Since UE $1$ is the weak user, power-domain NOMA requires $p_{1,1}\geq p_{1,0}$, which is equivalent to $\beta\in(0,0.5]$. For a given $\delta_1$, the power-allocation problem is given by
    \begin{subequations}
        \label{P1.2}
        \begin{align}
             \max_{\beta_1}&~R_{1,0} + R_{1,1}
            \label{prob:SR_2user_power} \\
            \rm{s.t.} &~R_{1,1} \ge R_1^{\min},R_{1,0} \ge R_0^{\min},\beta_1\in(0,0.5].
        \end{align}
    \end{subequations}
    The QoS constraints impose bounds on $\beta_1$. The wired-user constraint $R_{1,0} \ge R_0^{\min}$ yields
        \begin{equation}
        \label{min_alpha0}
        \beta_1\geq\beta^{\min}_1(\delta_1) 
        \triangleq \frac{(2^{R_0^{\min}} - 1) \sigma^2}{A_0 P_{\max}}.
    \end{equation}
    The wireless-user constraint $R_{1,1} \ge R_1^{\min}$ yields
    \begin{equation}
        \label{max_alpha0}
        \beta_1\leq\beta^{\max}_1(\delta_1) 
        \triangleq\frac{A_1 P_{\max} - (2^{R_1^{\min}} - 1) \sigma^2}{A_1 P_{\max}2^{R_1^{\min}}}.
    \end{equation}
Hence, the feasible set of $\beta_1$ is given by
   \begin{equation}
        \label{feasibleinterval_alpha}
        \beta_1 \in \left[\beta^{\min}_1(\delta_1),\min \{\beta^{\max}_1(\delta_1),0.5\}\right].
    \end{equation}

    Next, define $z \triangleq \frac{A_1 P_{\max}}{\sigma^2}$ and $m \triangleq \frac{A_0 P_{\max}}{\sigma^2}$. The corresponding SINR/SNR terms are $\gamma_{1,1}(\beta_1) = \frac{z(1-\beta_1)}{z \beta_1 + 1}$ and $\gamma_{1,0}(\beta_1) = m \beta_1$. The sum rate can be written as $R_{1,0} + R_{1,1}=\log_2(\frac{z+1}{z\beta_1 + 1}  (1+m\beta_1))$. Since $\log_2(\cdot)$ is strictly increasing, maximizing the sum rate is equivalent to maximizing $f(\beta_1) \triangleq \frac{z+1}{z \beta_1 + 1} (1 + m \beta_1)$. Its derivative is $f'(\beta_1) = (z+1)\frac{m-z}{(z\beta_1 + 1)^2}$. Under the NOMA-relevant condition $A_0\geq A_1$ (equivalently $m\geq z$), $f(\beta_1)$ is nondecreasing in $\beta_1$, and it is strictly increasing when $A_0> A_1$. Therefore, the optimal solution is attained at the largest feasible $\beta_1$, which yields
    \begin{equation}
        \label{alpha0}
        \beta_1^\star(\delta_1) = \min\{\beta^{\max}_1(\delta_1),0.5\}.
    \end{equation}
    Feasibility requires $\beta_1^\star(\delta_1)\ge\beta^{\min}_1(\delta_1)$. The resulting optimal power allocation is given by
    \begin{equation}
        \label{optimal_powerallocation01}
p_{1,0}^{\star} = \beta^{\star}_1(\delta_1) P_{\max},~ p_{1,1}^{\star} = (1-\beta^{\star}_1(\delta_1)) P_{\max}.
    \end{equation}
\subsubsection{Optimization of the PA Radiation}
This subsection examines how the PA radiation coefficient $\delta_1$ affects the feasible power-allocation interval and the resulting sum rate. Recall from \eqref{feasibleinterval_alpha} that the QoS constraints yield $\beta^{\min}_1(\delta_1)\leq \beta_1\leq\beta^{\max}_1(\delta_1)$, where both bounds depend on $\delta_1$ through $A_0$ and $A_1$. Using the chain rule, we have $\frac{{\rm{d}}\beta_1^{\max}(\delta_1)}{{\rm{d}}\delta_1} = \frac{{\rm{d}}\beta_1^{\max}}{{\rm{d}}A_1}\frac{{\rm{d}}A_1}{{\rm{d}}\delta_1} = \frac{(2^{R^{\min}_1}-1) \sigma^2 D_{\rm{v}}^2 {\rm{e}}^{2\alpha D_{\rm{i}}}}{2^{R^{\min}_1}\delta_1^3 P_{\max}} > 0$ and $\frac{{\rm{d}}\beta_1^{\min}(\delta_1)}{{\rm{d}}\delta_1} = \frac{{\rm{d}}\beta_1^{\min}}{{\rm{d}}A_0}\frac{{\rm{d}}A_0}{{\rm{d}}\delta_1} = \frac{(2^{R^{\min}_0}-1) \sigma^2 -{\rm{e}}^{2\alpha D_x}}{{\kappa_{\rm{c}}^2 (1-\delta_1)^2} P_{\max}} > 0$, where the strict positivity follows from $A_1$ increasing with $\delta_1$ and $A_0$ decreasing with $\delta_1$ under the adopted coupling model.
\vspace{-5pt}
    \begin{remark}
        \label{alpha0_delta1}
        Both $\beta^{\min}_1(\delta_1)$ and $\beta^{\max}_1(\delta_1)$ increase with $\delta_1$. Increasing $\delta_1$ therefore shifts the feasible interval of $\beta_1$ toward larger values. In addition, $\delta_1$ controls the relative strength of the radiated wireless link and the residual guided link. This effect changes the achievable rates and alters the feasible region through the QoS constraints and the SIC feasibility condition.
    \end{remark}
\vspace{-5pt}
Given the optimal PA position in \eqref{SS_Optimal_Location}, the SIC order in \eqref{optimalSIC}, and the optimal power allocation $\beta_1^\star(\delta_1)$ in \eqref{alpha0}, the remaining design variable is $\delta_1$. The resulting one-dimensional optimization is
    \begin{subequations}
        \label{onedimensionalsearch}
        \begin{align}
             \max_{\delta_1}&~R_{1,0} + R_{1,1}= \log_2\left(\frac{(z+1)(1+m\beta_1^\star(\delta_1))}{z\beta_1^\star(\delta_1) + 1} \right)\\
            \rm{s.t.} &~\delta_1\in(0,1),\\
            &~\log_2(1 + \gamma_{1,1}) \le \log_2(1+\gamma_{1,0 \rightarrow 1})\label{SIC_Feasibaility}.
        \end{align}
    \end{subequations}
    The SIC feasibility \eqref{SIC_Feasibaility} under the selected order requires that UE $0$ remains the strong user, which is guaranteed by $A_1\leq A_0$. Using \eqref{channelgain_11} and \eqref{channelgain_10}, this condition becomes  
\begin{align}
\frac{\eta\delta_1}{D_{\rm{v}}^2}{\rm{e}}^{-2\alpha D_{\rm{i}}}\leq \kappa_{\rm{c}}^2 (1-\delta_1){\rm{e}}^{-2\alpha D_{x}},
\end{align}  
which yields the following upper bound on $\delta_1$:
\begin{align}
\frac{1}{\delta_1}-1\geq\frac{\eta{\rm{e}}^{-2\alpha D_{\rm{i}}}}{D_{\rm{v}}^2\kappa_{\rm{c}}^2 {\rm{e}}^{-2\alpha D_{x}}}\Leftrightarrow
\delta_1\leq\frac{1}{1+\frac{\eta{\rm{e}}^{-2\alpha D_{\rm{i}}}}{D_{\rm{v}}^2\kappa_{\rm{c}}^2 {\rm{e}}^{-2\alpha D_{x}}}}\triangleq\delta_1^{\diamond}.
\end{align}
With this bound, the radiation design reduces to the following:
        \begin{align}\label{onedimensionalsearch1}
             \max_{\delta_1\in(0,\delta_1^{\diamond}]}~f_{\delta_1}(\delta_1)\triangleq \frac{(z+1)(1+m\beta_1^\star(\delta_1))}{z\beta_1^\star(\delta_1) + 1}.
        \end{align}
This problem is one-dimensional over a bounded interval, so it can be solved efficiently by a low-complexity line search. A simple implementation discretizes $(0,\delta_1^{\diamond}]$ into a $Q_{\delta_1}$-point grid, evaluates $f_{\delta_1}(\cdot)$ on the grid, and selects the best value. Let ${\mathcal{Q}}_{\delta_1}\triangleq \left\{0,\frac{\delta_1^{\diamond}}{Q_{\delta_1}-1},\frac{2\delta_1^{\diamond}}{Q_{\delta_1}-1},\ldots,\delta_1^{\diamond}\right\}$. A near-optimal solution to problem \eqref{onedimensionalsearch1} is then given by
\begin{align}\label{onedimensionalsearch2}
\delta_1^{\star}\triangleq\arg\max\nolimits_{x\in {\mathcal{Q}}_{\delta_1}}f_{\delta_1}(x).
\end{align}

The overall procedure for jointly optimizing the PA position, the PA radiation coefficient, and the power allocation in the two-user case is summarized in Algorithm \ref{alg:two_user}.}

\begin{algorithm}[!t]
    \caption{Optimal PA Position, Radiation, and Power Allocation of T-PASS for Two-User Case}
    \label{alg:two_user}
    
    \begin{algorithmic}[1]
    \REQUIRE $P_{\max},\sigma^2,R_0^{\min},R_1^{\min},\alpha,D_x$
    \ENSURE $\psi^{\star}_1,p_{1,0}^\star,p_{1,1}^\star,\delta_1^\star$
    \STATE Optimize the PA position by \eqref{SS_Optimal_Location}
    \STATE Optimize the PA radiation coefficient by \eqref{onedimensionalsearch2}
    \STATE Optimize the power allocation by \eqref{optimal_powerallocation01}
    \end{algorithmic}
    \end{algorithm}

\section{Multiuser Case}\label{Section: Multiuser Case}
\subsection{Problem Reformulation and Protocol Design}
{\color{black}We now consider the multiuser setting with $K$ wireless UEs. Following Remarks \ref{remark_noma_1} and \ref{remark_noma_2}, under typical PASS configurations the wired UE attains a much larger channel power gain than a wireless UE within the service region. We therefore treat UE 0 as the strong user in every H-NOMA slot and fix the SIC decoding order in slot $k\in[K]$ as follows:
    \begin{equation}
        \label{optimalSIC_multi}
        \Omega_k(k)=1,\quad\Omega_k(0)=2.
    \end{equation}
With this order, the achievable rates of wireless UE $k$ and UE $0$ in slot $k$ are given by
\begin{subequations}
\begin{align}
R_{k,k} &= \log_2\left(1 +  \frac{|h_{k,k}|^2 P_{\max}(1-\beta_k)}{|h_{k,k}|^2 P_{\max}\beta_k + \sigma^2}\right),\label{R_1_gain_multi}\\
    R_{k,0} &= \log_2\left(1+\frac{|h_{k,0}|^2 \beta_k P_{\max}}{\sigma^2}\right),\label{R_0_gain_multi}
\end{align}
\end{subequations}
where $\beta_k \triangleq \frac{p_{k,0}}{P_{\max}}$ denotes the power-allocation coefficient assigned to UE $0$ in slot $k$. The WSR maximization problem in \eqref{wsumrate_k} can then be rewritten as follows:
\begin{subequations}
    \label{wsumrate_k1}
    \begin{align}
    &\max_{\psi_{n}(\cdot),\delta_n(\cdot),\beta_k}
            \sum\nolimits_{k=1}^{K}\tau_k(w_{0}R_{k,0} + w_{k}R_{k,k})
        \label{prob:WSR_multiuser1} \\
        &\quad\qquad{\rm{s.t.}} ~  \tau_kR_{k,k} \ge R_1^{\min}, k\in[K],\label{c:R1min1}\\
        &\quad\qquad\quad~~ \sum\nolimits_{k=1}^{K} \tau_k R_{k,0} \ge R_0^{\min},\label{c:R0min1}\\
        &\quad\qquad\quad~~ |h_{k,0}|^2\geq |h_{k,k}|^2,k\in[K],
        \label{c:SIC_k1}\\
        &\quad\qquad\quad~~ \beta_k\in(0,0.5],k\in[K],
        \label{c:power1}\\
        &\quad\qquad\quad~~\sum\nolimits_{k=1}^{K} \tau_k =1, 0 \le \tau_k \le 1,k\in[K],\label{c:time_all1}\\
        & \quad\qquad\quad~~\psi_n(k) \in [0,D_x], k \in [K], n \in [N],
        \label{c:position1}\\
        & \quad\qquad\quad~~|\psi_n(k)-\psi_{n'}(k)|\geq \Delta_{\min}, n\ne n',
        \label{c:position11}\\
        & \quad\qquad\quad~~\delta_n(k)\in[0,1], k \in [K], n \in [N],
        \label{c:delta1}
    \end{align}
\end{subequations}
where \eqref{c:SIC_k1} enforces the strong-user condition associated with the fixed SIC order.

The performance of T-PASS depends on the reconfigurability of the PA radiation coefficients and PA positions along the waveguide. We therefore consider four operating protocols that reflect different hardware capabilities. The protocols differ in whether the PA positions $\{\psi_n(k)\}$ and radiation coefficients $\{\delta_n(k)\}$ are shared across the $K$ TDMA slots. \emph{Fixed-radiation fixed-position (FR-FP)} uses the same PA positions and the same radiation coefficients for all slots. \emph{Fixed-radiation adaptive-position (FR-AP)} uses the same radiation coefficients for all slots, while the PA positions are optimized per slot. \emph{Adaptive-radiation fixed-position (AR-FP)} uses the same PA positions for all slots, while the radiation coefficients are optimized per slot. \emph{Adaptive-radiation adaptive-position (AR-AP)} optimizes both PA positions and radiation coefficients in each slot. These protocols represent different levels of flexibility in configuring the PAs for the scheduled wireless UE in each time slot.}
\subsection{Algorithm Design}
{\color{black}
\subsubsection{{FR-FP}}
Under the FR-FP model, the PA radiation coefficients and PA positions are identical across all TDMA slots. The system therefore uses a single static PA configuration for all wireless UEs. Only the per-slot power allocation and the TDMA time-sharing factors are optimized. FR-FP has the lowest hardware complexity, but it provides the least flexibility in adapting the effective channel to different users. Let $\mathcal{C}_{\rm FRFP}$ denote the set of feasible PA configurations under FR-FP, which is defined as follows:
\begin{equation}
    \label{candidatePA_FRFP}
    \mathcal{C}_{\rm FRFP}\triangleq
\left\{(\{\psi_n(k)\},\{\delta_n(k)\})\left\lvert\begin{matrix}\psi_n(k)=\bar{\psi}_n,\forall n,k,\\
\delta_n(k)=\bar{\delta}_n,\forall n,k\end{matrix}\right.\right\},
\end{equation}
where $\bar{\psi}_n$ and $\bar{\delta}_n$ denote the slot-invariant PA position and radiation coefficient, respectively. 

For a fixed PA configuration $(\{\bar{\psi}_n\},\{\bar{\delta}_n\})$, the power-allocation coefficient $\beta_k$ controls the rate trade-off in slot $k$. The wired rate $R_{k,0}(\beta_k)$ increases with $\beta_k$ over the feasible interval, while the wireless rate $R_{k,k}(\beta_k)$ decreases with $\beta_k$. The QoS constraints and the NOMA power ordering define a feasible interval of the following form:
\begin{equation}
    \label{feasibleset_betak}
    \beta_k \in [\beta^{\min}_k, \min\{0.5, \beta_{k}^{\max}\}],
\end{equation}
where $\beta^{\min}_k$ is induced by the wired-user QoS constraint and $\beta_{k}^{\max}$ is induced by the wireless-user QoS constraint. The $\beta_{k}^{\max}$ and $\beta^{\min}_k$ can be expressed as $\beta_{k}^{\max} = \frac{1}{2^{R_1^{\min}}}\left(1-\frac{(2^{R_1^{\min}}-1)\sigma^2}{P_{\max}\lvert h_{k,k}\rvert^2}\right)$ and $\beta^{\min}_k = \large\frac{(2^{R_0^{\min}}-1)\sigma^2}{P_{\max}\lvert h_{k,0}\rvert^2}$.
Since the slot utility $w_{0}R_{k,0} + w_{k}R_{k,k}$ is maximized by pushing $\beta_k$ to the largest feasible value under the fixed SIC order, the optimal coefficient is given by:
\begin{equation}
    \label{optimal_betak}
    \beta_k^\star = \min\{0.5, \beta_{k}^{\max}\}.
\end{equation}
The corresponding optimal power allocation is
\begin{equation}
    \label{optimal_power_k}
    p_{k,0}^\star = \beta_k^\star P_{\max},\quad
    p_{k,k}^\star = \big(1-\beta_k^\star\big)P_{\max},\quad k\in [K].
\end{equation}

With $\{\beta_k^\star\}$ fixed and $(\{\bar{\psi}_n\},\{\bar{\delta}_n\})$ fixed, the per-slot weighted sum rate becomes a constant. The remaining optimization over the time-sharing factors $\{\tau_k\}$ is therefore a linear program. Define the minimum time fraction required to meet the wireless QoS in each slot $k\in[K]$ as $\tau_k^{\min}\triangleq \frac{R_1^{\min}}{R_{k,k}(\beta_k^\star)}$. Feasibility requires $\sum_{k=1}^{K} \tau_k^{\min} \le 1$ and $\sum_{k=1}^{K} \tau_k^{\min} R_{k,0}(\beta_k^\star) \ge R_0^{\min}$. The optimal time allocation assigns the remaining time budget to the slot that yields the largest weighted sum rate. Specifically, let $k^\star \triangleq \arg\max_{k\in[K]} (w_k R_{k,1}(\beta_k^\star) + w_0 R_{k,0}(\beta_k^\star))$  and $\Delta\tau \triangleq 1 - \sum_{k=1}^{K} \tau_k^{\min} \ge 0$. Then the optimal time-sharing factors take the following form:
\begin{equation}
    \label{optimal_time}
    \tau_k^\star =
    \begin{cases}
        \tau_k^{\min} + \Delta\tau, & k = k^\star, \\
        \tau_k^{\min},             & k \neq k^\star.
    \end{cases}
\end{equation}

The above discussion implies that, for any given $(\{\bar{\psi}_n\},\{\bar{\delta}_n\})$, the optimal power allocation and time-slot allocation can be computed in closed form. Substituting the optimal power allocation in \eqref{optimal_power_k} and the optimal time-sharing factors in \eqref{optimal_time} into the WSR objective in \eqref{prob:WSR_multiuser1}, the WSR maximization reduces to optimizing the PA positions $\bar{\bm\psi}\triangleq[\bar{\psi}_1,\ldots,\bar{\psi}_N]^{\mathsf{T}}$ and the PA radiation coefficients $\bar{\bm{\delta}}\triangleq[\bar{\delta}_1,\ldots,\bar{\delta}_N]^{\mathsf{T}}$ as follows:
\begin{subequations}
    \label{wsumrate_k1a}
    \begin{align}
    \max_{\bar{\bm\psi},\bar{\bm{\delta}}}&~f_{\rm FRFP}(\bar{\bm\psi},\bar{\bm{\delta}})
        \label{prob:WSR_multiuser1a} \\
        {\rm{s.t.}} &~  |h_{k,0}|^2\geq |h_{k,k}|^2,k\in[K],
        \label{c:SIC_k1a}\\
        & ~\bar{\psi}_n \in [0,D_x], n \in [N],
        \label{c:position1a}\\
        & ~|\bar{\psi}_n-\bar{\psi}_{n'}|\geq \Delta_{\min}, n\ne n',
        \label{c:position11a}\\
        & ~\bar{\delta}_n\in[0,1],n \in [N],
        \label{c:delta1a}
    \end{align}
\end{subequations}
where $f_{\rm FRFP}(\bar{\bm\psi},\bar{\bm{\delta}})$ denotes the resulting WSR after substituting \eqref{optimal_power_k} and \eqref{optimal_time}. 

Problem \eqref{wsumrate_k1a} is non-convex due to the tight coupling between $\bar{\bm\psi}$ and $\bar{\bm\delta}$. To handle this coupling, we adopt a block coordinate descent (BCD) procedure that partitions the decision variables into $2N$ scalar blocks $\{\bar{\psi}_n,\bar{\delta}_N\}_{n=1}^{N}$. Each block is updated in an alternating manner while the remaining variables are fixed, and the iterations continue until convergence. When updating $\bar{\psi}_n$ or $\bar{\delta}_n$, we perform a one-dimensional search over the corresponding interval $[0,D_x]$ or $[0,1]$, respectively. The candidate points are pruned to satisfy the SIC feasibility constraint in \eqref{c:SIC_k1a} and the minimum inter-PA spacing constraint in \eqref{c:position11a}. This yields an efficient element-wise BCD implementation with low per-iteration complexity.
\subsubsection{FR-AP}
Under the FR-AP protocol, the PA radiation coefficients are shared across slots, while the PA positions are adjusted to match the scheduled wireless UE in each slot. This protocol exploits geometric reconfigurability with a fixed radiation profile. Let $\mathcal{C}_{\rm FRAP}$ denote the set of feasible PA configurations under FR-AP, which is defined as follows:
\begin{equation}
    \label{candidatePA_FRAP}
    \mathcal{C}_{\rm FRAP}\triangleq
\left\{(\{\psi_n(k)\},\{\delta_n(k)\})\left\lvert\begin{matrix}\bar{\bm{\psi}}(k)\in{\mathcal{F}}_{\psi},\forall k,\\
\delta_n(k)=\bar{\delta}_n,\forall n,k\end{matrix}\right.\right\},
\end{equation}
where $\bar{\bm{\psi}}(k)\triangleq[\psi_1(k),\ldots,\psi_N(k)]^{\mathsf{T}}$ collects the PA positions used in slot $k$. The feasible PA-position set ${\mathcal{F}}_{\psi}$ is given by
\begin{align}
{\mathcal{F}}_{\psi}\triangleq\left\{[\psi_1,\ldots,\psi_N]^{\mathsf{T}}\left\lvert\begin{matrix}{\psi}_{n}\in[0,D_x],n\in[N],\\
\lvert{\psi}_{n}-{\psi}_{n'}\rvert\geq\Delta_{\min},n\ne n'\end{matrix}\right.\right\}.
\end{align}

For a fixed $\bar{\bm{\delta}}=[\bar{\delta}_1,\ldots,\bar{\delta}_N]^{\mathsf{T}}$, the WSR depends on $\{\bar{\bm{\psi}}(k)\}$ through the effective channel gains of the scheduled wireless UEs. The PA positions in each slot can therefore be designed to strengthen the radiated link of the scheduled wireless UE under the geometric constraints in ${\mathcal{F}}_{\psi}$. This per-slot PA-position design can be carried out using the PA position refinement method developed in our previous work \cite{ouyang2025array}. Let $\bar{\bm{\psi}}^{\star}(k)$ denote the resulting PA-position solution in slot $k$. 

With $\{\bar{\bm{\psi}}^{\star}(k)\}$ determined, the remaining variables follow the same structure as in FR-FP. The per-slot power-allocation coefficients are obtained from \eqref{optimal_power_k}, and the optimal time-sharing factors follow \eqref{optimal_time}. If desired, the common radiation coefficients ${\bar{\bm\delta}}$ can be further optimized via an element-wise BCD procedure, which applies the same one-dimensional radiation update used in the FR-FP design while keeping the per-slot PA positions fixed.

\subsubsection{AR-FP}
Under the AR-FP protocol, the PA positions are shared across all TDMA slots, while the PA radiation coefficients are tuned according to the scheduled wireless UE in each slot. This protocol keeps the mechanical configuration static and exploits only radiation reconfigurability. Let $\mathcal{C}_{\rm ARFP}$ denote the feasible set of PA configurations under AR-FP, which is defined as follows:
\begin{equation}
    \label{candidatePA_ARFP}
    \mathcal{C}_{\rm ARFP}\triangleq\left\{(\{\psi_n(k)\},\{\delta_n(k)\})\left\lvert\begin{matrix}\psi_n(k)=\bar{\psi}_n,\forall n,k,\\
\bar{\bm{\delta}}(k)\in{\mathcal{F}}_{\delta},\forall k\end{matrix}\right.\right\},
\end{equation}
where $\bar{\bm{\delta}}(k)\triangleq[\delta_1(k),\ldots,\delta_N(k)]^{\mathsf{T}}$ collects the radiation coefficients used in slot $k$. The feasible radiation set is given by ${\mathcal{F}}_{\delta}\triangleq\{[\delta_1,\ldots,\delta_N]^{\mathsf{T}}|\delta_n\in[0,1],n\in[N]\}$. 

Substituting the optimal power allocation in \eqref{optimal_power_k} and the optimal time-sharing factors in \eqref{optimal_time} into the WSR in \eqref{prob:WSR_multiuser1}, the WSR maximization reduces to optimizing the slot-invariant PA positions $\bar{\bm\psi}$ and the per-slot radiation vectors $\{\bar{\bm{\delta}}(k)\}$, which is given as follows:
\begin{subequations}
    \label{wsumrate_k1c}
    \begin{align}
    \max_{\bar{\bm\psi},\bar{\bm{\delta}}(k)}&~f_{\rm ARFP}(\bar{\bm\psi},\{\bar{\bm{\delta}}(k)\})
        \label{prob:WSR_multiuser1c} \\
        {\rm{s.t.}} &~  \eqref{c:SIC_k1a},\eqref{c:position1a},\eqref{c:position11a},\bar{\bm{\delta}}(k)\in{\mathcal{F}}_{\delta},\forall k,
        \label{c:delta1c}
    \end{align}
\end{subequations}
where $f_{\rm ARFP}(\bar{\bm\psi},\{\bar{\bm{\delta}}(k)\})$ denotes the resulting WSR after substituting \eqref{optimal_power_k} and \eqref{optimal_time}. The variables $\bar{\bm\psi}$ and $\{\bar{\bm{\delta}}(k)\}$ are coupled through the effective channel gains. We therefore adopt a BCD procedure that partitions the decision variables into $1+K$ blocks, namely $\{\bar{\bm\psi},\bar{\bm{\delta}}(1),\ldots,\bar{\bm{\delta}}(K)\}$. Each block is updated alternately while the remaining blocks are fixed, and the iterations continue until convergence. When updating $\bar{\bm\psi}$, we apply an element-wise one-dimensional search over $[0,D_x]$ for each $\bar{\psi}_n$. When updating $\bar{\bm{\delta}}(k)$, we apply an element-wise one-dimensional search over $[0,1]$ for each $\delta_n(k)$. In both updates, candidate points are pruned to satisfy the strong-user condition in \eqref{c:SIC_k1a} and the minimum inter-PA spacing constraint in \eqref{c:position11a}. This yields an efficient element-wise BCD implementation with low per-iteration complexity.
\subsubsection{AR-AP}
Under the AR-AP protocol, both the PA positions and the PA radiation coefficients are reconfigured in each TDMA slot. This protocol offers the highest flexibility and typically provides the best performance, at the cost of increased implementation complexity. Let $\mathcal{C}_{\rm ARAP}$ denote the feasible set of PA configurations under AR-AP, which is defined as follows:
\begin{equation}
    \label{candidatePA_ARAP}
    \mathcal{C}_{\rm ARAP}=\left\{(\{\psi_n(k)\},\{\delta_n(k)\})\left\lvert\begin{matrix}\bar{\bm{\psi}}(k)\in{\mathcal{F}}_{\psi},\forall k,\\
\bar{\bm{\delta}}(k)\in{\mathcal{F}}_{\delta},\forall k\end{matrix}\right.\right\}.
\end{equation}
In each slot $k$, the PA positions are chosen to strengthen the radiated link of the scheduled wireless UE under the geometric constraints in ${\mathcal{F}}_{\psi}$. This per-slot PA-position design is obtained using the PA position refinement method in \cite{ouyang2025array}. After determining the slot-wise PA positions, and substituting the closed-form power allocation in \eqref{optimal_power_k} and the optimal time-sharing factors in \eqref{optimal_time}, the remaining optimization reduces to selecting the radiation vectors $\{\bar{\bm{\delta}}(k)\}$, which is formulated as follows:
\begin{subequations}
    \label{wsumrate_k1d}
    \begin{align}
    \max_{\bar{\bm{\delta}}(k)}&~f_{\rm ARAP}(\{\bar{\bm{\delta}}(k)\})
        \label{prob:WSR_multiuser1d} \\
        {\rm{s.t.}} &~  |h_{k,0}|^2\geq |h_{k,k}|^2,k\in[K],
        \label{c:SIC_k1d}\\
        & ~{\delta}_n(k)\in[0,1],n \in [N],k\in[K],
        \label{c:delta1a}
    \end{align}
\end{subequations}
where $f_{\rm ARAP}(\bar{\bm\psi},\bar{\bm{\delta}})$ denotes the resulting WSR after substituting \eqref{optimal_power_k} and \eqref{optimal_time} and applying the PA-position refinement in \cite{ouyang2025array}. 

Problem \eqref{wsumrate_k1d} is still non-convex because the radiation coefficients affect both the wireless effective gain and the residual guided gain. We adopt a BCD procedure that partitions the decision variables into $K$ blocks $\{\bar{\bm{\delta}}(1),\ldots,\bar{\bm{\delta}}(K)\}$. Each block is updated alternately while the remaining blocks are fixed. Within each block update, an element-wise one-dimensional search over $[0,1]$ is applied to each $\delta_n(k)$, and infeasible candidate points are discarded to satisfy the SIC feasibility condition in \eqref{c:SIC_k1d}.}

\section{Numerical Results}\label{Section: Numerical Results}
{\color{black}We evaluate the proposed hybrid-NOMA T-PASS framework through numerical simulations. Unless stated otherwise, the carrier frequency is set to $f_c = 28$ GHz. The average in-waveguide attenuation factor is $\kappa = 0.08$ dB/m (equivalently, $\alpha=0.0092$ ${\text{m}}^{-1}$), the effective refractive index of the waveguide is set to $n_{\text{eff}} = 1.4$, and the impedance-matched waveguide-to-receiver coupler factor is set to $\kappa_{\text{c}}= 0.84$. The system bandwidth is $180$~MHz, the maximum transmit power is $P_{\max} = 20$ dBm, and the noise power is $\sigma^2 = -90$ dBm. A square service region of size $100 \times 100~\text{m}^2$ is considered, with waveguide length $D_x = 100$ m. A single dielectric waveguide is deployed along the $x$-axis at height $d = 3$ m, and the feed point is located at $[0,0,3]$ m. One wired UE is connected at the waveguide termination at $\mathbf{u}_0 = [100,0,3]$ m. The $K$ wireless UEs are independently and uniformly distributed over the service region, with $u^x_k \in[0,100]$ m, $u^y_k \in [-50,50]$ m. 

For the two-user case, one PA is activated, that is, $N=1$. For the multiuser case, the waveguide is equipped with $N=8$ PAs. The PAs are initialized at uniformly spaced positions and can slide within $[0,100]$ m. In the one-dimensional searches used in this arctic, the number of sampling points used to discretizing the corresponding interval is set to $10^4$. The minimum rate requirements are set to $R_0^{\min}  = R_1^{\min} = 1$ bit/s/Hz for the wired and wireless UEs. Unless otherwise specified, all curves are obtained by Monte Carlo averaging over $10^3$ independent realizations. In each realization, large-scale path loss and small-scale fading are generated according to the adopted channel model, and the proposed algorithms are executed to compute the resulting rates and WSR.}

\subsection{Two-User Case}

\begin{figure}[htbp]
    \centering
    \includegraphics[width=0.4\textwidth]{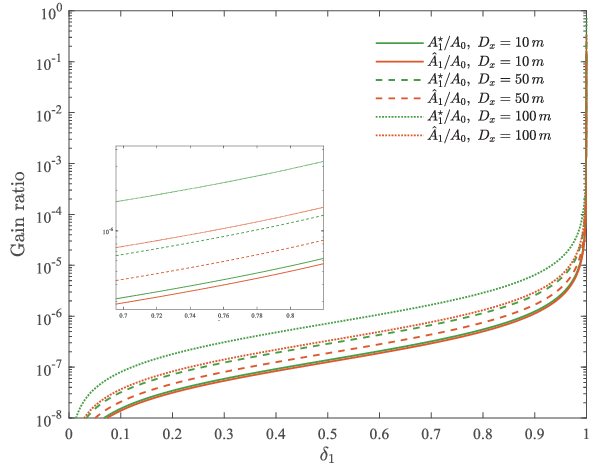}
    \caption{Gain ratios versus the radiation factor $\delta_1$ for the two-user case.}
    \label{fig_4}
\end{figure}

{\figurename} {\ref{fig_4}} illustrates the channel-gain ratios in \eqref{A0A1_maxratio} and \eqref{A1A0_ratio} versus the PA radiation coefficient $\delta_1$ for different waveguide lengths $D_x$. As $\delta_1$ increases from $0$ to approximately $0.6$, all curves exhibit a slow and nearly linear growth, indicating that the radiated wireless gain remains several orders of magnitude weaker than the residual guided gain. A sharp increase appears as $\delta_1$ becomes extremely close to one, where the radiated component starts to dominate. This behavior supports the physical interpretation that the guided link preserves a stronger effective channel unless almost all guided power is extracted for radiation. For all considered values of $D_x$, $\frac{A_1^\star}{A_0}$ consistently exceeds $\frac{\hat{A}_1}{A_0}$, which matches the analytical relationship between the idealized and averaged gain models. In addition, both ratios increase as $D_x$ increases. This trend follows from the exponential factor ${\rm{e}}^{2\alpha D_x}$, which increases the relative contribution of the radiated component with respect to the residual guided signal. Overall, the results indicate that, under typical settings, the wired UE generally experiences a much stronger effective channel than the wireless UE.

% When $\delta=1.0$ (all energy allocated to the wired UE), the proposed T-PASS still preserves at least $80\%$ of the rate of the conventional PASS for the wireless UE, while it is able to simultaneously serve an additional wired UE. As a result, the achievable sum rate of T-PASS is noticeably higher than that of the single-user conventional PASS. 
% As $\delta=0.7$, the rate of the wireless UE under T-PASS is no smaller than that of the conventional PASS with the same power splitting factor, and meanwhile the wired UE also enjoys a non-negligible data rate contribution, which further boosts the overall sum rate.
% Moreover, the converged rates clearly show that introducing the wired UE does not destabilize the optimization, since the residual guided signal is deterministically coupled with the PA radiation coefficient and can be efficiently accommodated by the derived closed-form updates.
% Therefore, these results confirm the effectiveness of the proposed T-PASS framework, which maintains comparable or even higher sum rate than the conventional PASS, while providing a significantly higher sum rate by supporting both wired and wireless users in the same framework.

\begin{figure}[htbp]
    \centering
    \includegraphics[width=0.4\textwidth]{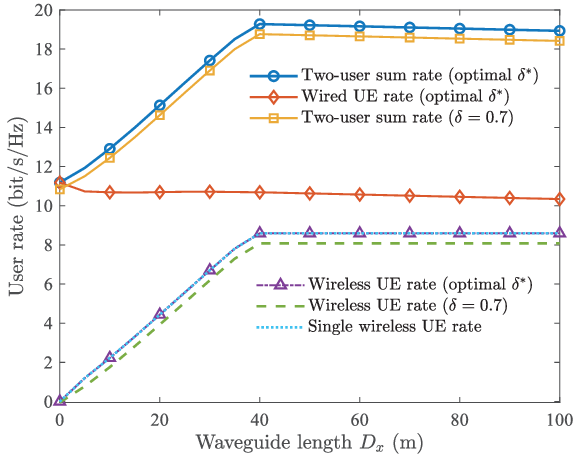}
    \caption{User rates of T-PASS versus waveguide length $D_x$ under different $\delta$ for the two-user case.}
    \label{fig_6}
\end{figure}

Fig.~\ref{fig_6} shows the impact of the waveguide length $D_x$ on the two-user T-PASS performance. With $\delta=0.7$, the wireless-UE rate under T-PASS remains no smaller than about 80$\%$ of that achieved by conventional PASS with the same power-splitting factor, while the wired UE contributes a non-negligible additional rate, which increases the sum rate. With the optimized $\delta^\star$, T-PASS preserves essentially the same wireless-UE rate as conventional PASS, while achieving a higher sum rate due to the added wired stream.
As $D_x$ increases, the sum rate under $\delta^\star$ rises rapidly at first, saturates around $D_x = 40$ m, and then decreases slightly for larger $D_x$. When $D_x$ increases from $20$ m to $40$ m, the PA can be positioned closer to the wireless UE with a more favorable radiation distance, which strengthens the wireless effective channel gain and improves both the wireless rate and the sum rate. When $D_x$ exceeds $40$ m, the placement benefit becomes marginal, while the guided signal experiences a longer propagation distance and stronger attenuation. This effect reduces the wired-link gain and slightly degrades the overall effective channels, which leads to a mild reduction in both the wired-UE rate and the sum rate.
Across all $D_x$, the optimized $\delta^\star$ yields a higher sum rate than the fixed $\delta=0.7$ setting, while keeping a comparable wireless-UE rate. The single-wireless-user baseline stays below the two-user curves over the entire range of $D_x$, which confirms that T-PASS supports simultaneous wired and wireless services with a substantially higher sum rate than conventional wireless-only PASS. The wired-UE rate also varies only mildly with $D_x$, which indicates that the residual guided link provides a reliable throughput contribution even when the PA position is adjusted to enhance the wireless link.

\begin{figure}[htbp]
    \centering
    \includegraphics[width=0.4\textwidth]{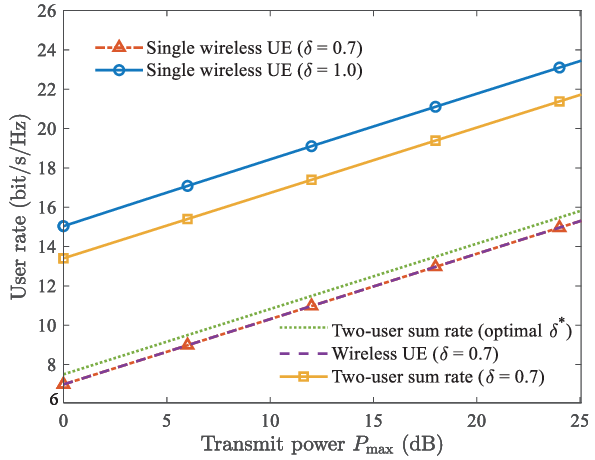}
    \caption{User rates of T-PASS versus transmit power $P_{\max}$ under different $\delta$ for the two-user case.}
    \label{fig_8}
\end{figure}

Fig.~\ref{fig_8} illustrates the impact of the transmit power $P_{\max}$ on the two-user case T-PASS performance. When $\delta=1.0$ (all energy allocated to the wired UE), the proposed T-PASS still preserves at least $80\%$ of the rate of the conventional PASS for the wireless UE, while it is able to simultaneously serve an additional wired UE.
With a fixed $\delta=0.7$, the wireless-UE rate under T-PASS almost coincides with the single-wireless-UE baseline using the same $\delta$ over the entire power range. This result indicates that T-PASS preserves essentially the same wireless performance as conventional PASS, even though it simultaneously serves an additional wired UE. Meanwhile, the two-user sum rate with $\delta= 0.7$ remains strictly above both single-UE baselines because the wired UE contributes additional throughput without a noticeable loss at the wireless UE.
When the radiation coefficient is further optimized to $\delta^{\star}$, the sum rate increases again and stays highest among all schemes for all $P_{\max}$. Therefore, T-PASS not only preserves the wireless-UE performance of conventional PASS, but it also achieves a substantial sum-rate improvement through the additional wired stream and the optimized $\delta^\star$. In this setting, the optimal $\delta^\star$ yields a 51$\%$ sum-rate gain relative to conventional PASS with $\delta= 0.7$.
These results also highlight the dual role of $\delta$ in T-PASS. It controls the power split between the radiated wireless link and the residual guided link, and it shapes the feasible power-allocation region through the QoS and SIC conditions under the fixed decoding order.

\subsection{Multiuser Case}

\begin{figure}[htbp]
    \centering
    \includegraphics[width=0.4\textwidth]{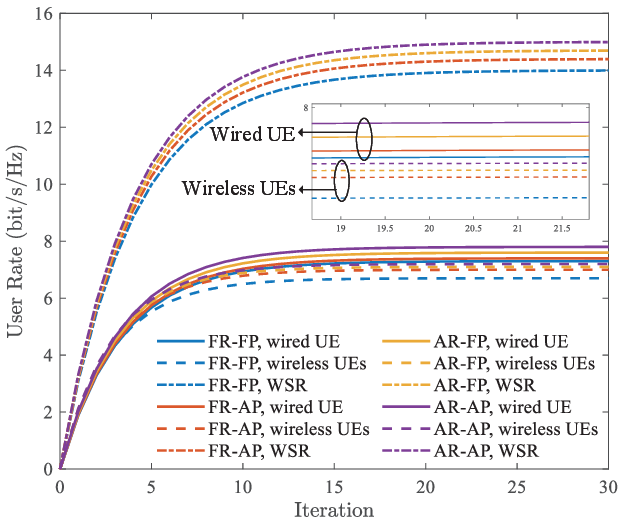}
    \caption{Convergence behavior of the proposed algorithms under four protocols for the multiuser case.}
    \label{fig_5}
\end{figure}

For the multiuser case, the WSR weights are fixed as $w_0 = \frac{5}{12}, w_k = \frac{7}{12}$ for the wired UE and all wireless UEs for the effective performance. The wireless UEs are assigned a slightly larger weight since the wired UE benefits from a more stable guided channel, whereas the wireless UEs rely on radiated links that are more sensitive to propagation conditions. Moreover, this setting enables effective performance balancing between the wired and wireless services \cite{yang2016weightNOMA}.

Fig.~\ref{fig_5} shows the convergence behaviors of the proposed algorithms under the four protocols FR-FP, FR-AP, AR-FP, and AR-AP for the multiuser case. All schemes quickly approach their steady-state values within about $10$ iterations of the iterative procedure, while their performance gaps are mainly determined by whether the PA radiation coefficients and PA positions are adaptive. FR-FP fixes both the radiation coefficient and PA locations, so the power split between wired and wireless links as well as the effective channel gains of wireless UEs cannot be adjusted, leading to the lowest wireless WSR and total WSR. FR-AP relaxes the radiation coefficient while keeping PA positions fixed. By better balancing the energy between the waveguide and the radiated links, it improves both the wireless WSR and the sum rate over FR-FP. AR-FP instead optimizes PA positions under a fixed radiation coefficient, which reduces the path loss to the wireless UEs and yields a similar or slightly higher gain. The fully adaptive AR-AP protocol jointly tunes both radiation and PA locations, thus exploiting all available degrees of freedom, and it achieves the highest WSR and also a slightly larger wired UE rate. Overall, Fig.~\ref{fig_5} shows that making the PA radiation and positions adaptive monotonically enhances the multiuser T-PASS performance, thereby exploiting the maximum degrees of freedom offered by the proposed T-PASS framework.

\begin{figure}[htbp]
    \centering
    \includegraphics[width=0.4\textwidth]{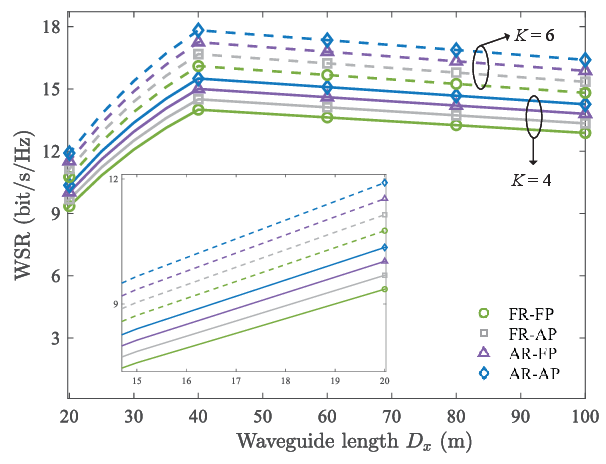}
    \caption{WSRs of T-PASS versus waveguide length $D_x$ under different protocols and user numbers for the multiuser case.}
    \label{fig_7}
\end{figure}

Fig.~\ref{fig_7} plots the WSR versus the waveguide length $D_x$ for different protocols and user numbers. For both $K=4$ and $K=6$, the WSR increases as $D_x$ grows from $20$ m to about $40$ m, and then saturates or decreases slightly for larger $D_x$. The initial increase comes from the added geometric flexibility provided by a longer waveguide, which enables better PA placement relative to the scheduled wireless UEs and strengthens the effective channel gains. When $D_x$ exceeds roughly $40$ m, the placement benefit becomes marginal, while the guided signal experiences a longer propagation distance and stronger in-waveguide attenuation, which limits further WSR improvement.
At any fixed $D_x$, the protocol ordering is consistent. AR-AP achieves the highest WSR, followed by AR-FP and FR-AP, while FR-FP yields the lowest WSR because neither the PA positions nor the radiation coefficients are adapted. The curves with $K = 6$ remain above those with $K = 4$ across all $D_x$, which reflects the multiuser diversity gain brought by serving more wireless UEs and accumulating more rate terms in the WSR. Overall, these results show that adaptivity in either PA positions or radiation coefficients yields clear gains, and jointly optimizing both provides the largest improvement.

\begin{figure}[htbp]
    \centering
    \includegraphics[width=0.4\textwidth]{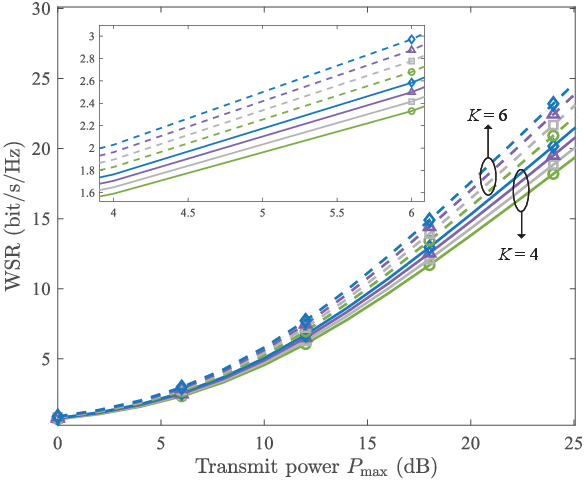}
    \caption{WSRs of T-PASS versus transmit power $P_{\max}$ under different protocols and user numbers for the multiuser case. The legend is the same as that of Fig. 8.}
    \label{fig_9}
\end{figure}

Fig.~\ref{fig_9} plots the WSR versus the transmit power $P_{\max}$ under different protocols and user numbers. For both $K=4$ and $K=6$, all curves increase almost linearly with $P_{\max}$, which matches the expected SINR improvement at higher transmit power. At any given $P_{\max}$, the protocol ordering remains unchanged across the whole power range, which indicates that the benefit of PA adaptivity is robust across different SNR regimes. FR-FP again provides the lowest WSR. Allowing a single degree of adaptivity in FR-AP or AR-FP yields a clear improvement, and AR-AP achieves the highest WSR by tuning both PA positions and radiation coefficients. The WSR curves with $K=6$ are uniformly above those with $K=4$, which further confirms the multiuser diversity advantage of the proposed T-PASS framework, including in the low power regime highlighted by the inset.

\section{Conclusion}\label{Section: Conclusion}

This paper proposed a hybrid-NOMA empowered T-PASS framework that integrates wired and wireless communications over a single dielectric waveguide. The framework jointly optimized PA positions, PA radiation coefficients, and power allocation to maximize system performance. In the two-user case, we established the optimal SIC decoding order analytically and showed that the wired UE acts as the strong user under typical T-PASS configurations. We also showed that the PA radiation coefficient plays a dual role: it shapes the channel power gains of both the wireless and wired links, and it directly governs the feasible interval of the power-allocation coefficient through the SIC and QoS constraints. In the multiuser case, we introduced four practical protocols, namely FR-FP, FR-AP, AR-FP, and AR-AP, which captures different levels of flexibility in configuring PA positions and radiation coefficients. Comparisons across these protocols showed that adaptivity in PA radiation and PA positioning improves the WSR, and that AR-AP achieves the highest WSR by jointly tuning both. Future work will extend the proposed framework to scenarios with multiple wired UEs and more general channel models, and will investigate robust designs under imperfect channel knowledge and practical PA actuation constraints.

\bibliographystyle{IEEEtran}
\def\baselinestretch{1}
\bibliography{TPASS}

\end{document}